\documentclass[pra,superscriptaddress,twocolumn]{revtex4-2}
\usepackage{enumerate}
\usepackage{amsfonts,amssymb,amsmath}
\usepackage[]{graphics,graphicx,epsfig}
\usepackage{amsthm}

\usepackage[colorlinks = true, citecolor=blue, linkcolor=blue, urlcolor=blue]{hyperref}
\newcommand*{\fullref}[1]{\hyperref[{#1}]{\autoref*{#1} \nameref*{#1}}}

\usepackage{graphicx}
\usepackage{dcolumn}
\usepackage{natbib}
\usepackage{color}
\usepackage{multirow}
\usepackage{ulem}

\begin{document} 

\title{Reexamination of the Kochen-Specker theorem: Relaxation of the completeness assumption}

\author{Kelvin Onggadinata}
\affiliation{Centre for Quantum Technologies, National University of Singapore, 3 Science Drive 2, 117543 Singapore, Singapore}
\affiliation{Department of Physics, National University of Singapore, 3 Science Drive 2, 117543 Singapore, Singapore}

\author{Dagomir Kaszlikowski}
\email{phykd@nus.edu.sg}
\affiliation{Centre for Quantum Technologies, National University of Singapore, 3 Science Drive 2, 117543 Singapore, Singapore}
\affiliation{Department of Physics, National University of Singapore, 3 Science Drive 2, 117543 Singapore, Singapore}

\author{Pawe{\l} Kurzy{\'n}ski}
\email{pawel.kurzynski@amu.edu.pl}
\affiliation{Institute of Spintronics and Quantum Information, Faculty of Physics, Adam Mickiewicz University, 61-614 Pozna\'n, Poland}
\affiliation{Centre for Quantum Technologies, National University of Singapore, 3 Science Drive 2, 117543 Singapore, Singapore}

\date{\today}


\begin{abstract}
The Kochen-Specker theorem states that exclusive and complete deterministic outcome assignments are impossible for certain sets of measurements, called Kochen-Specker (KS) sets. A straightforward consequence is that KS sets do not have joint probability distributions because no set of joint outcomes over such a distribution can be constructed. However, we show it is possible to construct a joint quasiprobability distribution over any KS set by relaxing the completeness assumption. Interestingly, completeness is still observable at the level of measurable marginal probability distributions. This suggests the observable completeness might not be a
fundamental feature, but a secondary property.

\end{abstract}

\maketitle


\section{Introduction}

Violations of Bell inequalities \cite{bell1964einstein,brunner2014nonlocality,giustina2015significant,shalm2015strong,hensen2016loophole}, or non-contextuality inequalities \cite{cabello2008experimentally,budroni2022kochen,kirchmair2009state,wang2022significant}, imply a lack of a joint probability distribution (JPD) over a set of corresponding measurements \cite{fine1982hidden,liang2011specker}. Let us consider one of the simplest examples: the Wright/Klyachko-Can-Binicioglu-Shumovsky (Wright/KCBS) inequality \cite{wright1978mathematical,klyachko2008simple}
\begin{equation}\label{KCBS}
\sum_{i=1}^5 \langle A_i \rangle \leq 2.
\end{equation}
It involves five events to which one assigns a binary $\{0,1\}$ random variable ({\it measurement}) $\{A_i\}_{i=1}^{5}$. The events are cyclically exclusive, i.e., if $A_i = 1$, then $A_{i\pm 1} = 0$ (summing is $\text{mod}\, 5$). Moreover, these events are cyclically co-measurable, meaning $A_i$ can be jointly measured with $A_{i\pm 1}$, but not with $A_{i\pm 2}$. The Wright/KCBS scenario can be implemented on a quantum three-level system (qutrit), in which case $\{A_i\}_{i=1}^{5}$ are cyclically orthogonal projective rank one measurements. If the qutrit is in a maximally mixed state $\rho = \openone/3$, then $\text{Tr}\{\rho A_i\} = 1/3$ for all $i$ and the inequality (\ref{KCBS}) is not violated. In this case there exists a classical JPD
\begin{equation}
p(A_1=a_1,\ldots,A_5=a_5) \equiv p(a_1,\ldots,a_5), 
\end{equation}
where $a_i \in \{0,1\}$. It recovers all measurable marginal probabilities $p(a_i,a_{i\pm 1})$. Such a JPD is not unique so here is an example: 
\begin{eqnarray}
p(1,0,1,0,0) &=& 1/6, \nonumber \\ 
p(1,0,0,1,0) &=& 1/6, \nonumber \\ 
p(0,1,0,1,0) &=& 1/6, \nonumber \\ 
p(0,1,0,0,1) &=& 1/6, \nonumber \\ 
p(0,0,1,0,1) &=& 1/6, \nonumber \\ 
p(0,0,0,0,0) &=& 1/6.
\end{eqnarray}
This JPD obeys the exclusivity relations, i.e., no two jointly measurable properties are both assigned the value of 1. However, there exists a set of measurements and a qutrit state $|\psi\rangle$ such that $\langle \psi | A_i| \psi \rangle = 1/\sqrt{5}$ for all $i$ \cite{klyachko2008simple}.  These measurements  violate (\ref{KCBS}) up to $\sqrt{5}$, excluding a positive JPD emulation. However, a quasi-probability distribution with negative probabilities (JQD) is possible \cite{abramsky2011sheaf,abramsky2014operational,alsafi2013simulating}, for instance:
\begin{eqnarray}
q(1,0,1,0,0) &=& 1/2\sqrt{5}, \nonumber \\ 
q(1,0,0,1,0) &=& 1/2\sqrt{5}, \nonumber \\ 
q(0,1,0,1,0) &=& 1/2\sqrt{5}, \nonumber \\ 
q(0,1,0,0,1) &=& 1/2\sqrt{5}, \nonumber \\ 
q(0,0,1,0,1) &=& 1/2\sqrt{5}, \nonumber \\ 
q(0,0,0,0,0) &=& 1 - 5/2\sqrt{5} \approx -0.118
\end{eqnarray}
does the job. It satisfies the exclusivity relations and recovers the measurable marginal probability distributions.

Although seemingly exotic, JQD is a well defined mathematical concept \cite{bartlett1945negative,szekely2005half,khrennikov2013non}, extensively used in quantum theory since Wigner function discovery \cite{feynman1987negative,ferrie2009framed,ferrie2011quasi,wigner1932quantum}. Recently, we demonstrated that JQDs can also be used as a computational resource to reach a nonclassical computing speedup \cite{kaszlikowski2021little}.  In addition, the JQD's negativity can be used as a measure of nonclassicality (``quantumness'') \cite{oas2014exploring,morris2022witnessing,onggadinata2021local}, hence the Wright-KCBS scenario classifies the maximally mixed state as classical and $|\psi\rangle$ as non-classical. 

Curiously, there are measurement scenarios, contextual for any quantum state \cite{cabello2008experimentally}, called state-independent contextuality (SI-C) \cite{budroni2022kochen}. Can one construct a JQD for any SI-C scenario? A positive answer to this question was given in \cite{abramsky2011sheaf} using this sheaf-theoretic approach. Here we provide a simple JQD construction for specific SI-C scenarios corresponding to Kochen-Specker (KS) sets and later we show that this construction works for arbitrary measurement scenarios, contextual or noncontextual. In addition, we focus on a particular of the origin of such JQDs. Similarly to JPDs, JQDs assign quasi-probabilities to all measurement events. Each such event corresponds to an outcome assignment to all observables at once. The flagship specimen is the KS theorem \cite{kochen1967problem}. It states that, for certain measurement sets, KS sets, it is impossible to find outcome assignments, satisfying {\it exclusivity} and {\it completeness}. Exclusivity means that no two measurement events can be observed at the same time. However, the completeness of a mutually exclusive event sets means that exactly one of these events will be observed. Formally it is as follows
\begin{enumerate}
\item {\it Exclusivity.} For a jointly measurable subset of mutually exclusive events, corresponding to $\{A_1,A_2\ldots,A_m\}$, at most one of them will occur at the same time, i.e., only the following outcome assignments $\{a_1,a_2,\ldots,a_m\}$ are allowed: $\{0,0,\ldots,0\}$, $\{1,0,\ldots,0\}$, $\{0,1,\ldots,0\}$, $\ldots$, $\{0,0,\ldots,1\}$.
\item {\it Completeness.} For a complete jointly measurable subset of mutually exclusive events, corresponding to $\{A_1,A_2\ldots,A_n\}$, exactly one of them will occur, i.e., only the following outcome assignments $\{a_1,a_2,\ldots,a_n\}$ are allowed: $\{1,0,\ldots,0\}$, $\{0,1,\ldots,0\}$, $\ldots$, $\{0,0,\ldots,1\}$.
\end{enumerate}

For the projective quantum measurements, the mutual exclusivity of projector subsets $S_E$ is imposed by their mutual orthogonality, i.e., $A_i A_j = \delta_{i,j} A_i$ for all pairs $\{A_i,A_j\}\in S_E$. However, a mutually exclusive subset of projectors $S_C$ is complete if $\sum_{A_i \in S_C} A_i = \openone$. Finally, note that any complete subset is exclusive and any exclusive subset can be extended to a complete subset $S_E \subseteq S_C$.

The exclusivity is a necessary ingredient of all contextuality scenarios, both state-dependent and state-independent. However, as far as we know, the completeness assumption is necessary for all known SI-C scenarios. In particular, all known KS sets contain complete subsets. Here we show that it is possible to construct a JQD for any KS set if one relaxes the completeness assumption. Moreover, our constructions are compatible with the quantum theory. These JQDs can be used to model realistic measurements on KS sets, where completeness is observed in the measurable marginal distributions. This strongly suggests that completeness might be a secondary property, rather than a fundamental phenomenon.

We also note that not all SI-C scenarios correspond to KS sets \cite{yu2012state}. In such cases outcome assignments are possible and therefore we can construct a JQD without relaxation of the completeness assumption. Nevertheless, later we are going to show that our method allows to find an alternative JQD for such scenarios. In fact, we show that it allows to find a JQD for an arbitrary measurement scenario.


\section{Quasiprobability distributions}

Before we show how to construct a JQD for a given KS set, let us present a simple idea of how observable completeness, as well as exclusivity, emerge in quasiprobability theories. 

Consider two events $A$ and $B$ with attached indicator random variables (indicators), $R_A$ and $R_B$. These indicators have outcome of 1 if the corresponding event occurs and 0 otherwise. Let us assume the events can be jointly measured and the indicators return one of the following outcomes $\{00,01,10,11\}$, where the first outcome corresponds to $R_A$ and the second one to $R_B$. A general probability distribution over these outcomes reads ${\mathbf p} = \{p_{00},p_{01},p_{10},p_{11}\}$. We do not assume $A$ and $B$'s exclusivity and completeness, so, in general, $p_{11}\neq 0$ and $p_{00}\neq 0$.

Next, consider a third event $C$ with the corresponding indicator $R_C$. Let us first assume that all three events are jointly measurable, hence a measurement returns one of the eight outcomes $\{000,001,\ldots,111\}$, where the last position corresponds to $R_C$. The corresponding probability distribution reads ${\mathbf q}=\{q_{000},q_{001},\ldots,q_{111}\}$. If we do not make any assumptions about exclusivity and completeness, the only constraint on $\mathbf q$ is 
\begin{equation}\label{constr}
q_{000}+q_{001}+\ldots+q_{111}=1. 
\end{equation}

Now, assume $\mathbf q$ is a quasi-probability distribution, i.e., some probabilities are negative, but still sum up to 1 as in (\ref{constr}). To exclude negative probabilities in the laboratory (we do not know what they mean), we postulate that $A$, $B$ and $C$ cannot be measured together (only $A$ and $B$ are comeasurable). In addition, we demand the marginal distribution over $A$ and $B$ to be positive
\begin{eqnarray}
p_{00} &=& q_{000} + q_{001} \geq 0, \nonumber \\
p_{01} &=& q_{010} + q_{011} \geq 0, \nonumber \\
p_{10} &=& q_{100} + q_{101} \geq 0, \nonumber \\
p_{11} &=& q_{110} + q_{111} \geq 0.
\end{eqnarray}
Remarkably, if $q_{111} = - q_{110}$, $A$ and $B$ become exclusive. In addition, if $q_{000} = - q_{001}$, we guarantee observable completeness. This shows that observable exclusivity and completeness are not fundamental but a secondary property.

The above quasiprobability scenario with three questions, where only two can be asked simultaneously, has been studied since the so-called Specker's triangle discovery \cite{liang2011specker,specker1960logik}. In particular, the Specker's triangle scenario assumes that one can measure jointly either $A$ and $B$, or $A$ and $C$, or $B$ and $C$. The corresponding quasi-probability distribution can be given by
\begin{eqnarray}
q^{(ST)}_{010} &=& q^{(ST)}_{100} = q^{(ST)}_{110} =  q^{(ST)}_{001} = q^{(ST)}_{011} = q^{(ST)}_{101} =\frac{1}{4}, \nonumber \\
q^{(ST)}_{000} &=& q^{(ST)}_{111} = -\frac{1}{4}.
\end{eqnarray} 
This distribution says that whatever two questions you ask, you always find that either one or the other occurs, each with probability $1/2$. For example, if one measures $A$ and $B$, the corresponding marginal distribution is
\begin{eqnarray}
p_{00} &=& q_{000} + q_{001} = 0, \nonumber \\
p_{01} &=& q_{010} + q_{011} = 1/2, \nonumber \\
p_{10} &=& q_{100} + q_{101} = 1/2, \nonumber \\
p_{11} &=& q_{110} + q_{111} = 0.
\end{eqnarray}
Therefore, the Specker's triangle exhibits both the exclusivity and completeness.

Let us explain in more detail what do we mean by the completeness and the exclusivity as a secondary property of some system. The fact that for two events, $A$ and $B$, one does not observe outcome $00$ and $11$ may simply be a fundamental property, namely, that the system has only two states corresponding to the outcomes $\{01, 10\}$. On the other hand, the system might have four states corresponding to $\{00,01,10,11\}$, but one simply prepares it in the anticorrelated state of $A$ and $B$, i.e., in the state $01$ with probability $p$ and $10$ with probability $1-p$. In this case the observable secondary completeness and exclusivity are trivial. However, in the case of the Specker's triangle (and in other contextuality scenarios) the situation is more complicated. One cannot assume the fundamental completeness and exclusivity, since there is not set of outcomes for $A$, $B$, and $C$ that would reduce to $\{01,10\}$ for each pair. In addition, one cannot assume the trivial secondary completeness and exclusivity, since there is no JPD over the set $\{000,001,\ldots,111\}$ that would explain the statistics of the Specker's triangle. However, as presented above, one can construct a JQD over $\{000,001, \ldots, 111\}$ that would explain the observable statistics and would give rise to the completeness and exclusivity. 


\section{JQD construction}

Here we propose a JQD for an arbitrary KS set. The KS set consists of $N$ events, corresponding to $\{A_1,A_2,\ldots,A_N\}$. The smallest known KS set that can be implemented within quantum theory consists of $N=18$ events and requires a four-level quantum system \cite{cabello1996bell}. There are subsets of the KS set, commonly known as {\it measurement contexts}, that corresponds to jointly measurable sets of exclusive events. Some contexts are complete subsets (recall definitions above). Each KS set consists a proof of the KS theorem, i.e., there are no outcome assignments $\{a_1,a_2,\ldots,a_N\}$, where $a_i \in \{0,1\}$, such that for each complete context $\mathcal{C}_c$ one gets $\sum_{i\in\mathcal{C}_c}a_i = 1$. 

Quantum realization of a KS set consists of $N$ rank one projectors. Each measurement context consists of mutually orthogonal projectors and each complete context $\mathcal{C}_c$ has projectors such that $\sum_{i\in \mathcal{C}_c}A_i = \openone$. In particular, for rank-one projectors, the number of projectors in the complete set equals to the dimension of the Hilbert space of the corresponding quantum system.

Our JQD construction starts with an arbitrary preparation of the system that assigns to each event from the KS set a probability of its occurrence
\begin{equation}
p_i \equiv p(A_i = 1) \geq 0. 
\end{equation}
For a quantum system, we start with an arbitrary state $\rho$ that assigns a probability $p_i \equiv \text{Tr}\{\rho A_i\}$. Next, we give up the completeness assumption and allow for $N+1$ outcome assignments to all the events in the KS set:
\begin{equation}\label{eq: outcome notation 1}
\omega_i \equiv \{\underbrace{0,\ldots,0}_{i-1}1\underbrace{0,\ldots,0}_{N-i}\},
\end{equation} 
where $i=1,\ldots,N$, and
\begin{equation}\label{eq: outcome notation 2}
\omega_0 \equiv \{\underbrace{0,\ldots,0}_{N}\}.
\end{equation}
We assign to each $\omega_i$ ($i\neq 0$) the probability
\begin{equation}
p(\omega_i) \equiv p_i, ~~~~i\neq 0.
\end{equation}
Finally, we assign to $\omega_0$ the following quasiprobability
\begin{equation}
p(\omega_0) \equiv p_0 \equiv 1-\sum_{i=1}^N p_i < 0. 
\end{equation}
Note that $p_0$ is negative since for each complete measurement context $\mathcal{C}_c$ that is strictly included in the KS set the following holds:
\begin{equation}\label{cc}
\sum_{i\in \mathcal{C}_c} p_i =1,
\end{equation} 
therefore
\begin{equation}
1-\sum_{i=1}^N p_i < 1 - \sum_{i\in {\mathcal C}_c} p_i = 0. 
\end{equation}

Let us show that the above construction recovers observable marginal probability distributions for all measurement contexts, including complete ones. Consider a context $\mathcal{C}$ corresponding to an $n$-element subset $\{A_{1}^{({\mathcal C})},A_{2}^{({\mathcal C})},\ldots,A_{n}^{({\mathcal C})}\}$, where $n<N$. Each $A_{i}^{({\mathcal C})}$ ties to a different element $A_{j}$ from the KS set. The corresponding probability assignments are
\begin{equation}
p_i^{({\mathcal C})} = p(A_{i}^{({\mathcal C})}),~~~~i\neq 0,
\end{equation}
and the following holds
\begin{equation}
\sum_{i=1}^n p_i^{({\mathcal C})} \leq 1.
\end{equation}
The probability that none of the events in the context $\mathcal C$ occurs is 
\begin{equation}
p_0^{({\mathcal C})} = \left(\sum_{i=0}^N p_i\right) - \left(\sum_{j=1}^n p_j^{({\mathcal C})}\right) = 1 - \left(\sum_{j=1}^n p_j^{({\mathcal C})}\right) \geq 0.
\end{equation}
Finally, if $\mathcal C$ is a complete context, i.e., ${\mathcal C}= {\mathcal C}_c$, then (\ref{cc}) holds and $p_0^{({\mathcal C}_c)} =0$. Therefore, the completeness is recovered at the level of marginals.

It is important to notice that in our JQD construction the relaxation of the completeness assumptions turns a KS set into a non-KS set. Nevertheless, the quasiprobabilities are assigned such that measurable marginal distributions obey completeness assumptions. Therefore, at the level of allowed observations, the system described by our JQD cannot be distinguished from any system described by the corresponding KS set, hence our JQD provides a description of such a system.


\section{Discussion}

Although we discussed KS sets, the proposed JQD construction applies to an arbitrary set of measurement events $\{A_1, A_2, \ldots ,A_M\}$. For such a set we can define outcome assignments $\{\omega_0,\omega_1, \ldots , \omega_M\}$ in the same way as we did in Eqs. \eqref{eq: outcome notation 1} and \eqref{eq: outcome notation 2}. These outcome assignments may require relaxation of the completeness assumption, even if the corresponding set of measurements is not a KS set. The probabilities corresponding to $\{\omega_0, \ldots , \omega_M\}$ stem from a particular preparation of the system (e.g., $p_i\equiv \text{Tr}\{\rho A_i\}$ in a quantum case) and the quasiprobability corresponding to $\omega_0$ is $p_0 \equiv 1-\sum_{i=1}^M p_i$. What is important, due to the arguments presented in the previous section, such a JQD recovers all completeness relations at the level of observable marginal distributions. Finally, note that this JQD is nonunique, i.e., there might be some other JQD, perhaps over a different set of outcome assignments, that exhibits less negativity, or even no negativity, in which case it becomes a JPD and the corresponding scenario is clearly noncontextual. However, in some cases our construction may also lead to a JPD. In particular, if the set of measurements and the preparation of the system yield $\sum_{i=1}^M p_i \leq 1$, then $p_0 \geq 0$.

\begin{figure}[!htb]
    \centering
    \includegraphics[width=0.9\linewidth]{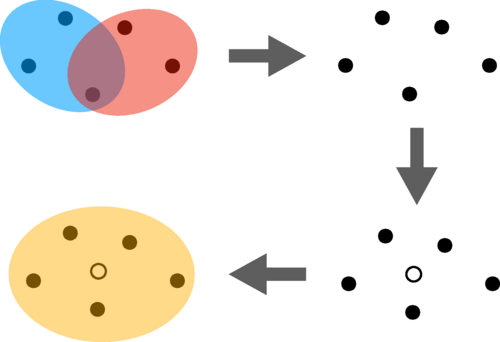}
    \caption{Representation of our idea using the hypergraph approach. Above we present an example hypergraph with fiver vertices and two hyperedges corresponding to two complete sets, each having three elements. These two sets overlap. Relaxation of the completeness assumptions is visualized as an erasure of the hyperedges. The new blank vertex represents the outcome assignment $\omega_0$ that corresponds to a situation in which none of the previous events happens. Finally, assigning this vertex with a quasiprobability allows us to introduce a new hyperedge. See text for more details.}
    \label{fig: jqd construction}
\end{figure}

To get a better understanding of our idea, it is worth using a graph representation in which measurement events $\{A_1,A_2,\ldots, A_M\}$ are depicted as vertices $V(G)=\{v_1,v_2,\ldots , v_M\}$ of some graph $G$ \cite{amaral2018graph}. There are two approaches. One can consider exclusivity graphs \cite{cabello2014graph}, in which case the edges $E(G)$ represent exclusivity relations between the events, i.e., if $A_i$ and $A_j$ are exclusive, then $\{v_iv_j\}\in E(G)$. Moreover, an edge $\{v_i,v_j\}$ implies that the respective probabilities $p_i + p_j \leq 1$. Alternatively, one can consider hypergraphs \cite{acin2015combinatorial}, in which case hyperedges $H(G)$ represent complete subsets, i.e., if $\{A_i,A_j,\ldots,A_k\}$ forms a complete subsets, then $\{v_i,v_j,\ldots , v_k\}\in H(G)$. Such a hyperedge implies that the respective probabilities obey $p_i+p_j + \ldots + p_k=1$. Here we use the hypergraph approach to represent our idea, see Fig. \ref{fig: jqd construction}. By introducing outcome assignments $\{\omega_1,\omega_2,\ldots,\omega_M\}$ we drop the completeness assumptions, which can be visualized as the erasure of all hyperedges (the first step in Fig. \ref{fig: jqd construction}). Next, we introduce the outcome assignment $\omega_0$, which can be visualized as the addition of an auxiliary vertex (blank vertex, the second step in Fig. \ref{fig: jqd construction}) to the graph. This vertex can be interpreted as a complementary event that, together with the previous events, forms a new complete set which can be assigned a new hyperedge (the third step in Fig. \ref{fig: jqd construction}). This new hyperedge implies $p_0+p_1+\ldots +p_M=1$ since $\omega_0$ is assigned the quasiprobability $1-\sum_{i=1}^M p_i$.

Let us also mention that our JQD construction applies to a continuous set of measurements. Note that the precursor of the KS theorem, the Gleason's theorem \cite{gleason1957measures} that is about a continuous sets of projective measurements, also relies on exclusivity and completeness. Our approach allows to relax the Gleason's assumptions and to assign quasiprobabilities over a continuous set.

An additional consequence of our result is that JQDs allow for a unified nonclassicality measure of both, state-dependent and state-independent contextuality scenarios. Of course, since we did not prove that our construction is optimal in a sense that the corresponding negativity is minimal, the nonclassicality measure based on our JQDs may overestimate the nonclassicality of KS sets. Nevertheless, we believe that our method is a step on the way of establishing such measures.


\section*{Acknowledgement}

D.K. is supported by the National Research Foundation, Singapore, and A*Star under the CQT Bridging Grant. P.K. is supported by the Polish National Science Centre (NCN) under the Maestro Grant no. DEC-2019/34/A/ST2/00081.

\bibliography{references}

\begin{thebibliography}{35}%
\makeatletter
\providecommand \@ifxundefined [1]{%
 \@ifx{#1\undefined}
}%
\providecommand \@ifnum [1]{%
 \ifnum #1\expandafter \@firstoftwo
 \else \expandafter \@secondoftwo
 \fi
}%
\providecommand \@ifx [1]{%
 \ifx #1\expandafter \@firstoftwo
 \else \expandafter \@secondoftwo
 \fi
}%
\providecommand \natexlab [1]{#1}%
\providecommand \enquote  [1]{``#1''}%
\providecommand \bibnamefont  [1]{#1}%
\providecommand \bibfnamefont [1]{#1}%
\providecommand \citenamefont [1]{#1}%
\providecommand \href@noop [0]{\@secondoftwo}%
\providecommand \href [0]{\begingroup \@sanitize@url \@href}%
\providecommand \@href[1]{\@@startlink{#1}\@@href}%
\providecommand \@@href[1]{\endgroup#1\@@endlink}%
\providecommand \@sanitize@url [0]{\catcode `\\12\catcode `\$12\catcode
  `\&12\catcode `\#12\catcode `\^12\catcode `\_12\catcode `\%12\relax}%
\providecommand \@@startlink[1]{}%
\providecommand \@@endlink[0]{}%
\providecommand \url  [0]{\begingroup\@sanitize@url \@url }%
\providecommand \@url [1]{\endgroup\@href {#1}{\urlprefix }}%
\providecommand \urlprefix  [0]{URL }%
\providecommand \Eprint [0]{\href }%
\providecommand \doibase [0]{https://doi.org/}%
\providecommand \selectlanguage [0]{\@gobble}%
\providecommand \bibinfo  [0]{\@secondoftwo}%
\providecommand \bibfield  [0]{\@secondoftwo}%
\providecommand \translation [1]{[#1]}%
\providecommand \BibitemOpen [0]{}%
\providecommand \bibitemStop [0]{}%
\providecommand \bibitemNoStop [0]{.\EOS\space}%
\providecommand \EOS [0]{\spacefactor3000\relax}%
\providecommand \BibitemShut  [1]{\csname bibitem#1\endcsname}%
\let\auto@bib@innerbib\@empty
\bibitem [{\citenamefont {Bell}(1964)}]{bell1964einstein}%
  \BibitemOpen
  \bibfield  {author} {\bibinfo {author} {\bibfnamefont {J.~S.}\ \bibnamefont
  {Bell}},\ }\href {https://doi.org/10.1103/PhysicsPhysiqueFizika.1.195}
  {\bibfield  {journal} {\bibinfo  {journal} {Phys. Phys. Fiz.}\ }\textbf
  {\bibinfo {volume} {1}},\ \bibinfo {pages} {195} (\bibinfo {year}
  {1964})}\BibitemShut {NoStop}%
\bibitem [{\citenamefont {Brunner}\ \emph {et~al.}(2014)\citenamefont
  {Brunner}, \citenamefont {Cavalcanti}, \citenamefont {Pironio}, \citenamefont
  {Scarani},\ and\ \citenamefont {Wehner}}]{brunner2014nonlocality}%
  \BibitemOpen
  \bibfield  {author} {\bibinfo {author} {\bibfnamefont {N.}~\bibnamefont
  {Brunner}}, \bibinfo {author} {\bibfnamefont {D.}~\bibnamefont {Cavalcanti}},
  \bibinfo {author} {\bibfnamefont {S.}~\bibnamefont {Pironio}}, \bibinfo
  {author} {\bibfnamefont {V.}~\bibnamefont {Scarani}},\ and\ \bibinfo {author}
  {\bibfnamefont {S.}~\bibnamefont {Wehner}},\ }\href
  {https://doi.org/10.1103/RevModPhys.86.419} {\bibfield  {journal} {\bibinfo
  {journal} {Rev. Mod. Phys.}\ }\textbf {\bibinfo {volume} {86}},\ \bibinfo
  {pages} {419} (\bibinfo {year} {2014})}\BibitemShut {NoStop}%
\bibitem [{\citenamefont {Giustina}\ \emph {et~al.}(2015)\citenamefont
  {Giustina}, \citenamefont {Versteegh}, \citenamefont {Wengerowsky},
  \citenamefont {Handsteiner}, \citenamefont {Hochrainer}, \citenamefont
  {Phelan}, \citenamefont {Steinlechner}, \citenamefont {Kofler}, \citenamefont
  {Larsson}, \citenamefont {Abell\'an}, \citenamefont {Amaya}, \citenamefont
  {Pruneri}, \citenamefont {Mitchell}, \citenamefont {Beyer}, \citenamefont
  {Gerrits}, \citenamefont {Lita}, \citenamefont {Shalm}, \citenamefont {Nam},
  \citenamefont {Scheidl}, \citenamefont {Ursin}, \citenamefont {Wittmann},\
  and\ \citenamefont {Zeilinger}}]{giustina2015significant}%
  \BibitemOpen
  \bibfield  {author} {\bibinfo {author} {\bibfnamefont {M.}~\bibnamefont
  {Giustina}}, \bibinfo {author} {\bibfnamefont {M.~A.~M.}\ \bibnamefont
  {Versteegh}}, \bibinfo {author} {\bibfnamefont {S.}~\bibnamefont
  {Wengerowsky}}, \bibinfo {author} {\bibfnamefont {J.}~\bibnamefont
  {Handsteiner}}, \bibinfo {author} {\bibfnamefont {A.}~\bibnamefont
  {Hochrainer}}, \bibinfo {author} {\bibfnamefont {K.}~\bibnamefont {Phelan}},
  \bibinfo {author} {\bibfnamefont {F.}~\bibnamefont {Steinlechner}}, \bibinfo
  {author} {\bibfnamefont {J.}~\bibnamefont {Kofler}}, \bibinfo {author}
  {\bibfnamefont {J.-A.}\ \bibnamefont {Larsson}}, \bibinfo {author}
  {\bibfnamefont {C.}~\bibnamefont {Abell\'an}}, \bibinfo {author}
  {\bibfnamefont {W.}~\bibnamefont {Amaya}}, \bibinfo {author} {\bibfnamefont
  {V.}~\bibnamefont {Pruneri}}, \bibinfo {author} {\bibfnamefont {M.~W.}\
  \bibnamefont {Mitchell}}, \bibinfo {author} {\bibfnamefont {J.}~\bibnamefont
  {Beyer}}, \bibinfo {author} {\bibfnamefont {T.}~\bibnamefont {Gerrits}},
  \bibinfo {author} {\bibfnamefont {A.~E.}\ \bibnamefont {Lita}}, \bibinfo
  {author} {\bibfnamefont {L.~K.}\ \bibnamefont {Shalm}}, \bibinfo {author}
  {\bibfnamefont {S.~W.}\ \bibnamefont {Nam}}, \bibinfo {author} {\bibfnamefont
  {T.}~\bibnamefont {Scheidl}}, \bibinfo {author} {\bibfnamefont
  {R.}~\bibnamefont {Ursin}}, \bibinfo {author} {\bibfnamefont
  {B.}~\bibnamefont {Wittmann}},\ and\ \bibinfo {author} {\bibfnamefont
  {A.}~\bibnamefont {Zeilinger}},\ }\href
  {https://doi.org/10.1103/PhysRevLett.115.250401} {\bibfield  {journal}
  {\bibinfo  {journal} {Phys. Rev. Lett.}\ }\textbf {\bibinfo {volume} {115}},\
  \bibinfo {pages} {250401} (\bibinfo {year} {2015})}\BibitemShut {NoStop}%
\bibitem [{\citenamefont {Shalm}\ \emph {et~al.}(2015)\citenamefont {Shalm},
  \citenamefont {Meyer-Scott}, \citenamefont {Christensen}, \citenamefont
  {Bierhorst}, \citenamefont {Wayne}, \citenamefont {Stevens}, \citenamefont
  {Gerrits}, \citenamefont {Glancy}, \citenamefont {Hamel}, \citenamefont
  {Allman}, \citenamefont {Coakley}, \citenamefont {Dyer}, \citenamefont
  {Hodge}, \citenamefont {Lita}, \citenamefont {Verma}, \citenamefont
  {Lambrocco}, \citenamefont {Tortorici}, \citenamefont {Migdall},
  \citenamefont {Zhang}, \citenamefont {Kumor}, \citenamefont {Farr},
  \citenamefont {Marsili}, \citenamefont {Shaw}, \citenamefont {Stern},
  \citenamefont {Abell\'an}, \citenamefont {Amaya}, \citenamefont {Pruneri},
  \citenamefont {Jennewein}, \citenamefont {Mitchell}, \citenamefont {Kwiat},
  \citenamefont {Bienfang}, \citenamefont {Mirin}, \citenamefont {Knill},\ and\
  \citenamefont {Nam}}]{shalm2015strong}%
  \BibitemOpen
  \bibfield  {author} {\bibinfo {author} {\bibfnamefont {L.~K.}\ \bibnamefont
  {Shalm}}, \bibinfo {author} {\bibfnamefont {E.}~\bibnamefont {Meyer-Scott}},
  \bibinfo {author} {\bibfnamefont {B.~G.}\ \bibnamefont {Christensen}},
  \bibinfo {author} {\bibfnamefont {P.}~\bibnamefont {Bierhorst}}, \bibinfo
  {author} {\bibfnamefont {M.~A.}\ \bibnamefont {Wayne}}, \bibinfo {author}
  {\bibfnamefont {M.~J.}\ \bibnamefont {Stevens}}, \bibinfo {author}
  {\bibfnamefont {T.}~\bibnamefont {Gerrits}}, \bibinfo {author} {\bibfnamefont
  {S.}~\bibnamefont {Glancy}}, \bibinfo {author} {\bibfnamefont {D.~R.}\
  \bibnamefont {Hamel}}, \bibinfo {author} {\bibfnamefont {M.~S.}\ \bibnamefont
  {Allman}}, \bibinfo {author} {\bibfnamefont {K.~J.}\ \bibnamefont {Coakley}},
  \bibinfo {author} {\bibfnamefont {S.~D.}\ \bibnamefont {Dyer}}, \bibinfo
  {author} {\bibfnamefont {C.}~\bibnamefont {Hodge}}, \bibinfo {author}
  {\bibfnamefont {A.~E.}\ \bibnamefont {Lita}}, \bibinfo {author}
  {\bibfnamefont {V.~B.}\ \bibnamefont {Verma}}, \bibinfo {author}
  {\bibfnamefont {C.}~\bibnamefont {Lambrocco}}, \bibinfo {author}
  {\bibfnamefont {E.}~\bibnamefont {Tortorici}}, \bibinfo {author}
  {\bibfnamefont {A.~L.}\ \bibnamefont {Migdall}}, \bibinfo {author}
  {\bibfnamefont {Y.}~\bibnamefont {Zhang}}, \bibinfo {author} {\bibfnamefont
  {D.~R.}\ \bibnamefont {Kumor}}, \bibinfo {author} {\bibfnamefont {W.~H.}\
  \bibnamefont {Farr}}, \bibinfo {author} {\bibfnamefont {F.}~\bibnamefont
  {Marsili}}, \bibinfo {author} {\bibfnamefont {M.~D.}\ \bibnamefont {Shaw}},
  \bibinfo {author} {\bibfnamefont {J.~A.}\ \bibnamefont {Stern}}, \bibinfo
  {author} {\bibfnamefont {C.}~\bibnamefont {Abell\'an}}, \bibinfo {author}
  {\bibfnamefont {W.}~\bibnamefont {Amaya}}, \bibinfo {author} {\bibfnamefont
  {V.}~\bibnamefont {Pruneri}}, \bibinfo {author} {\bibfnamefont
  {T.}~\bibnamefont {Jennewein}}, \bibinfo {author} {\bibfnamefont {M.~W.}\
  \bibnamefont {Mitchell}}, \bibinfo {author} {\bibfnamefont {P.~G.}\
  \bibnamefont {Kwiat}}, \bibinfo {author} {\bibfnamefont {J.~C.}\ \bibnamefont
  {Bienfang}}, \bibinfo {author} {\bibfnamefont {R.~P.}\ \bibnamefont {Mirin}},
  \bibinfo {author} {\bibfnamefont {E.}~\bibnamefont {Knill}},\ and\ \bibinfo
  {author} {\bibfnamefont {S.~W.}\ \bibnamefont {Nam}},\ }\href
  {https://doi.org/10.1103/PhysRevLett.115.250402} {\bibfield  {journal}
  {\bibinfo  {journal} {Phys. Rev. Lett.}\ }\textbf {\bibinfo {volume} {115}},\
  \bibinfo {pages} {250402} (\bibinfo {year} {2015})}\BibitemShut {NoStop}%
\bibitem [{\citenamefont {Hensen}\ \emph {et~al.}(2016)\citenamefont {Hensen},
  \citenamefont {Kalb}, \citenamefont {Blok}, \citenamefont {Dr{\'e}au},
  \citenamefont {Reiserer}, \citenamefont {Vermeulen}, \citenamefont
  {Schouten}, \citenamefont {Markham}, \citenamefont {Twitchen}, \citenamefont
  {Goodenough} \emph {et~al.}}]{hensen2016loophole}%
  \BibitemOpen
  \bibfield  {author} {\bibinfo {author} {\bibfnamefont {B.}~\bibnamefont
  {Hensen}}, \bibinfo {author} {\bibfnamefont {N.}~\bibnamefont {Kalb}},
  \bibinfo {author} {\bibfnamefont {M.}~\bibnamefont {Blok}}, \bibinfo {author}
  {\bibfnamefont {A.}~\bibnamefont {Dr{\'e}au}}, \bibinfo {author}
  {\bibfnamefont {A.}~\bibnamefont {Reiserer}}, \bibinfo {author}
  {\bibfnamefont {R.}~\bibnamefont {Vermeulen}}, \bibinfo {author}
  {\bibfnamefont {R.}~\bibnamefont {Schouten}}, \bibinfo {author}
  {\bibfnamefont {M.}~\bibnamefont {Markham}}, \bibinfo {author} {\bibfnamefont
  {D.}~\bibnamefont {Twitchen}}, \bibinfo {author} {\bibfnamefont
  {K.}~\bibnamefont {Goodenough}}, \emph {et~al.},\ }\href
  {https://doi.org/https://doi.org/10.1038/srep30289} {\bibfield  {journal}
  {\bibinfo  {journal} {Sci. Rep.}\ }\textbf {\bibinfo {volume} {6}},\ \bibinfo
  {pages} {30289} (\bibinfo {year} {2016})}\BibitemShut {NoStop}%
\bibitem [{\citenamefont {Cabello}(2008)}]{cabello2008experimentally}%
  \BibitemOpen
  \bibfield  {author} {\bibinfo {author} {\bibfnamefont {A.}~\bibnamefont
  {Cabello}},\ }\href {https://doi.org/10.1103/PhysRevLett.101.210401}
  {\bibfield  {journal} {\bibinfo  {journal} {Phys. Rev. Lett.}\ }\textbf
  {\bibinfo {volume} {101}},\ \bibinfo {pages} {210401} (\bibinfo {year}
  {2008})}\BibitemShut {NoStop}%
\bibitem [{\citenamefont {Budroni}\ \emph {et~al.}(2022)\citenamefont
  {Budroni}, \citenamefont {Cabello}, \citenamefont {G\"uhne}, \citenamefont
  {Kleinmann},\ and\ \citenamefont {Larsson}}]{budroni2022kochen}%
  \BibitemOpen
  \bibfield  {author} {\bibinfo {author} {\bibfnamefont {C.}~\bibnamefont
  {Budroni}}, \bibinfo {author} {\bibfnamefont {A.}~\bibnamefont {Cabello}},
  \bibinfo {author} {\bibfnamefont {O.}~\bibnamefont {G\"uhne}}, \bibinfo
  {author} {\bibfnamefont {M.}~\bibnamefont {Kleinmann}},\ and\ \bibinfo
  {author} {\bibfnamefont {J.-A.}\ \bibnamefont {Larsson}},\ }\href
  {https://doi.org/10.1103/RevModPhys.94.045007} {\bibfield  {journal}
  {\bibinfo  {journal} {Rev. Mod. Phys.}\ }\textbf {\bibinfo {volume} {94}},\
  \bibinfo {pages} {045007} (\bibinfo {year} {2022})}\BibitemShut {NoStop}%
\bibitem [{\citenamefont {Kirchmair}\ \emph {et~al.}(2009)\citenamefont
  {Kirchmair}, \citenamefont {Z{\"a}hringer}, \citenamefont {Gerritsma},
  \citenamefont {Kleinmann}, \citenamefont {G{\"u}hne}, \citenamefont
  {Cabello}, \citenamefont {Blatt},\ and\ \citenamefont
  {Roos}}]{kirchmair2009state}%
  \BibitemOpen
  \bibfield  {author} {\bibinfo {author} {\bibfnamefont {G.}~\bibnamefont
  {Kirchmair}}, \bibinfo {author} {\bibfnamefont {F.}~\bibnamefont
  {Z{\"a}hringer}}, \bibinfo {author} {\bibfnamefont {R.}~\bibnamefont
  {Gerritsma}}, \bibinfo {author} {\bibfnamefont {M.}~\bibnamefont
  {Kleinmann}}, \bibinfo {author} {\bibfnamefont {O.}~\bibnamefont
  {G{\"u}hne}}, \bibinfo {author} {\bibfnamefont {A.}~\bibnamefont {Cabello}},
  \bibinfo {author} {\bibfnamefont {R.}~\bibnamefont {Blatt}},\ and\ \bibinfo
  {author} {\bibfnamefont {C.~F.}\ \bibnamefont {Roos}},\ }\href
  {https://doi.org/https://doi.org/10.1038/nature08172} {\bibfield  {journal}
  {\bibinfo  {journal} {Nature}\ }\textbf {\bibinfo {volume} {460}},\ \bibinfo
  {pages} {494} (\bibinfo {year} {2009})}\BibitemShut {NoStop}%
\bibitem [{\citenamefont {Wang}\ \emph {et~al.}(2022)\citenamefont {Wang},
  \citenamefont {Zhang}, \citenamefont {Luan}, \citenamefont {Um},
  \citenamefont {Wang}, \citenamefont {Qiao}, \citenamefont {Xie},
  \citenamefont {Zhang}, \citenamefont {Cabello},\ and\ \citenamefont
  {Kim}}]{wang2022significant}%
  \BibitemOpen
  \bibfield  {author} {\bibinfo {author} {\bibfnamefont {P.}~\bibnamefont
  {Wang}}, \bibinfo {author} {\bibfnamefont {J.}~\bibnamefont {Zhang}},
  \bibinfo {author} {\bibfnamefont {C.-Y.}\ \bibnamefont {Luan}}, \bibinfo
  {author} {\bibfnamefont {M.}~\bibnamefont {Um}}, \bibinfo {author}
  {\bibfnamefont {Y.}~\bibnamefont {Wang}}, \bibinfo {author} {\bibfnamefont
  {M.}~\bibnamefont {Qiao}}, \bibinfo {author} {\bibfnamefont {T.}~\bibnamefont
  {Xie}}, \bibinfo {author} {\bibfnamefont {J.-N.}\ \bibnamefont {Zhang}},
  \bibinfo {author} {\bibfnamefont {A.}~\bibnamefont {Cabello}},\ and\ \bibinfo
  {author} {\bibfnamefont {K.}~\bibnamefont {Kim}},\ }\href
  {https://doi.org/https://doi.org/10.1126/sciadv.abk1660} {\bibfield
  {journal} {\bibinfo  {journal} {Sci. Adv.}\ }\textbf {\bibinfo {volume}
  {8}},\ \bibinfo {pages} {eabk1660} (\bibinfo {year} {2022})}\BibitemShut
  {NoStop}%
\bibitem [{\citenamefont {Fine}(1982)}]{fine1982hidden}%
  \BibitemOpen
  \bibfield  {author} {\bibinfo {author} {\bibfnamefont {A.}~\bibnamefont
  {Fine}},\ }\href {https://doi.org/10.1103/PhysRevLett.48.291} {\bibfield
  {journal} {\bibinfo  {journal} {Phys. Rev. Lett.}\ }\textbf {\bibinfo
  {volume} {48}},\ \bibinfo {pages} {291} (\bibinfo {year} {1982})}\BibitemShut
  {NoStop}%
\bibitem [{\citenamefont {Liang}\ \emph {et~al.}(2011)\citenamefont {Liang},
  \citenamefont {Spekkens},\ and\ \citenamefont {Wiseman}}]{liang2011specker}%
  \BibitemOpen
  \bibfield  {author} {\bibinfo {author} {\bibfnamefont {Y.-C.}\ \bibnamefont
  {Liang}}, \bibinfo {author} {\bibfnamefont {R.~W.}\ \bibnamefont
  {Spekkens}},\ and\ \bibinfo {author} {\bibfnamefont {H.~M.}\ \bibnamefont
  {Wiseman}},\ }\href
  {https://doi.org/https://doi.org/10.1016/j.physrep.2011.05.001} {\bibfield
  {journal} {\bibinfo  {journal} {Phys. Rep.}\ }\textbf {\bibinfo {volume}
  {506}},\ \bibinfo {pages} {1} (\bibinfo {year} {2011})}\BibitemShut {NoStop}%
\bibitem [{\citenamefont {Wright}(1978)}]{wright1978mathematical}%
  \BibitemOpen
  \bibfield  {author} {\bibinfo {author} {\bibfnamefont {R.}~\bibnamefont
  {Wright}},\ }\bibinfo {title} {in {M}athematical {F}oundations of {Q}uantum
  {T}heory}\ (\bibinfo  {publisher} {Academic Press},\ \bibinfo {address} {New
  York},\ \bibinfo {year} {1978})\ pp.\ \bibinfo {pages} {255--274}\BibitemShut
  {NoStop}%
\bibitem [{\citenamefont {Klyachko}\ \emph {et~al.}(2008)\citenamefont
  {Klyachko}, \citenamefont {Can}, \citenamefont
  {Binicio\ifmmode~\breve{g}\else \u{g}\fi{}lu},\ and\ \citenamefont
  {Shumovsky}}]{klyachko2008simple}%
  \BibitemOpen
  \bibfield  {author} {\bibinfo {author} {\bibfnamefont {A.~A.}\ \bibnamefont
  {Klyachko}}, \bibinfo {author} {\bibfnamefont {M.~A.}\ \bibnamefont {Can}},
  \bibinfo {author} {\bibfnamefont {S.}~\bibnamefont
  {Binicio\ifmmode~\breve{g}\else \u{g}\fi{}lu}},\ and\ \bibinfo {author}
  {\bibfnamefont {A.~S.}\ \bibnamefont {Shumovsky}},\ }\href
  {https://doi.org/10.1103/PhysRevLett.101.020403} {\bibfield  {journal}
  {\bibinfo  {journal} {Phys. Rev. Lett.}\ }\textbf {\bibinfo {volume} {101}},\
  \bibinfo {pages} {020403} (\bibinfo {year} {2008})}\BibitemShut {NoStop}%
\bibitem [{\citenamefont {Abramsky}\ and\ \citenamefont
  {Brandenburger}(2011)}]{abramsky2011sheaf}%
  \BibitemOpen
  \bibfield  {author} {\bibinfo {author} {\bibfnamefont {S.}~\bibnamefont
  {Abramsky}}\ and\ \bibinfo {author} {\bibfnamefont {A.}~\bibnamefont
  {Brandenburger}},\ }\href {https://doi.org/10.1088/1367-2630/13/11/113036}
  {\bibfield  {journal} {\bibinfo  {journal} {New J. Phys.}\ }\textbf {\bibinfo
  {volume} {13}},\ \bibinfo {pages} {113036} (\bibinfo {year}
  {2011})}\BibitemShut {NoStop}%
\bibitem [{\citenamefont {Abramsky}\ and\ \citenamefont
  {Brandenburger}(2014)}]{abramsky2014operational}%
  \BibitemOpen
  \bibfield  {author} {\bibinfo {author} {\bibfnamefont {S.}~\bibnamefont
  {Abramsky}}\ and\ \bibinfo {author} {\bibfnamefont {A.}~\bibnamefont
  {Brandenburger}},\ }\bibinfo {title} {An {O}perational {I}nterpretation of
  {N}egative {P}robabilities and {N}o-{S}ignalling {M}odels, in {H}orizons of
  the {M}ind. {A} {T}ribute to {P}rakash {P}anangaden, edited by {F}. van
  {B}reugel, {E}. {K}ashefi, {C}. {P}alamidessi, and {J}. {R}utten}\ (\bibinfo
  {publisher} {Springer},\ \bibinfo {address} {Cham},\ \bibinfo {year} {2014})\
  pp.\ \bibinfo {pages} {59--75}\BibitemShut {NoStop}%
\bibitem [{\citenamefont {Al-Safi}\ and\ \citenamefont
  {Short}(2013)}]{alsafi2013simulating}%
  \BibitemOpen
  \bibfield  {author} {\bibinfo {author} {\bibfnamefont {S.~W.}\ \bibnamefont
  {Al-Safi}}\ and\ \bibinfo {author} {\bibfnamefont {A.~J.}\ \bibnamefont
  {Short}},\ }\href {https://doi.org/10.1103/PhysRevLett.111.170403} {\bibfield
   {journal} {\bibinfo  {journal} {Phys. Rev. Lett.}\ }\textbf {\bibinfo
  {volume} {111}},\ \bibinfo {pages} {170403} (\bibinfo {year}
  {2013})}\BibitemShut {NoStop}%
\bibitem [{\citenamefont {Bartlett}(1945)}]{bartlett1945negative}%
  \BibitemOpen
  \bibfield  {author} {\bibinfo {author} {\bibfnamefont {M.~S.}\ \bibnamefont
  {Bartlett}},\ }\href {https://doi.org/10.1017/S0305004100022398} {\bibfield
  {journal} {\bibinfo  {journal} {Math. Proc. Camb. Philos. Soc.}\ }\textbf
  {\bibinfo {volume} {41}},\ \bibinfo {pages} {71–73} (\bibinfo {year}
  {1945})}\BibitemShut {NoStop}%
\bibitem [{\citenamefont {Sz{\'e}kely}(2005)}]{szekely2005half}%
  \BibitemOpen
  \bibfield  {author} {\bibinfo {author} {\bibfnamefont {G.~J.}\ \bibnamefont
  {Sz{\'e}kely}},\ }\href
  {https://web.archive.org/web/20131108010759/http://www.wilmott.com/pdfs/100609_gjs.pdf}
  {\bibfield  {journal} {\bibinfo  {journal} {Wilmott Magazine}\ }\textbf
  {\bibinfo {volume} {50}},\ \bibinfo {pages} {66} (\bibinfo {year}
  {2005})}\BibitemShut {NoStop}%
\bibitem [{\citenamefont {Khrennikov}(2013)}]{khrennikov2013non}%
  \BibitemOpen
  \bibfield  {author} {\bibinfo {author} {\bibfnamefont {A.~Y.}\ \bibnamefont
  {Khrennikov}},\ }\bibinfo {title} {Non-{A}rchimedean {A}nalysis: {Q}uantum
  {P}aradoxes, {D}ynamical {S}ystems and {B}iological {M}odels}\ (\bibinfo
  {publisher} {Springer Science \& Business Media},\ \bibinfo {year}
  {2013})\BibitemShut {NoStop}%
\bibitem [{\citenamefont {Feynman}(1987)}]{feynman1987negative}%
  \BibitemOpen
  \bibfield  {author} {\bibinfo {author} {\bibfnamefont {R.~P.}\ \bibnamefont
  {Feynman}},\ }\bibinfo {title} {Negative {P}robability, in {Q}uantum
  {I}mplication: {E}ssays in {H}onour of {D}avid {B}ohm, edited by {B}. {H}iley
  and {F}. {P}eat}\ (\bibinfo  {publisher} {Routledge},\ \bibinfo {address}
  {London},\ \bibinfo {year} {1987})\ pp.\ \bibinfo {pages}
  {235--248}\BibitemShut {NoStop}%
\bibitem [{\citenamefont {Ferrie}\ and\ \citenamefont
  {Emerson}(2009)}]{ferrie2009framed}%
  \BibitemOpen
  \bibfield  {author} {\bibinfo {author} {\bibfnamefont {C.}~\bibnamefont
  {Ferrie}}\ and\ \bibinfo {author} {\bibfnamefont {J.}~\bibnamefont
  {Emerson}},\ }\href {https://doi.org/10.1088/1367-2630/11/6/063040}
  {\bibfield  {journal} {\bibinfo  {journal} {New J. Phys.}\ }\textbf {\bibinfo
  {volume} {11}},\ \bibinfo {pages} {063040} (\bibinfo {year}
  {2009})}\BibitemShut {NoStop}%
\bibitem [{\citenamefont {Ferrie}(2011)}]{ferrie2011quasi}%
  \BibitemOpen
  \bibfield  {author} {\bibinfo {author} {\bibfnamefont {C.}~\bibnamefont
  {Ferrie}},\ }\href {https://doi.org/10.1088/0034-4885/74/11/116001}
  {\bibfield  {journal} {\bibinfo  {journal} {Rep. Prog. Phys.}\ }\textbf
  {\bibinfo {volume} {74}},\ \bibinfo {pages} {116001} (\bibinfo {year}
  {2011})}\BibitemShut {NoStop}%
\bibitem [{\citenamefont {Wigner}(1932)}]{wigner1932quantum}%
  \BibitemOpen
  \bibfield  {author} {\bibinfo {author} {\bibfnamefont {E.}~\bibnamefont
  {Wigner}},\ }\href {https://doi.org/10.1103/PhysRev.40.749} {\bibfield
  {journal} {\bibinfo  {journal} {Phys. Rev.}\ }\textbf {\bibinfo {volume}
  {40}},\ \bibinfo {pages} {749} (\bibinfo {year} {1932})}\BibitemShut
  {NoStop}%
\bibitem [{\citenamefont {Kaszlikowski}\ and\ \citenamefont
  {Kurzy{\'n}ski}(2021)}]{kaszlikowski2021little}%
  \BibitemOpen
  \bibfield  {author} {\bibinfo {author} {\bibfnamefont {D.}~\bibnamefont
  {Kaszlikowski}}\ and\ \bibinfo {author} {\bibfnamefont {P.}~\bibnamefont
  {Kurzy{\'n}ski}},\ }\href {https://doi.org/10.1007/s10701-021-00461-w}
  {\bibfield  {journal} {\bibinfo  {journal} {Found. Phys.}\ }\textbf {\bibinfo
  {volume} {51}} (\bibinfo {year} {2021})}\BibitemShut {NoStop}%
\bibitem [{\citenamefont {Oas}\ \emph {et~al.}(2014)\citenamefont {Oas},
  \citenamefont {de~Barros},\ and\ \citenamefont
  {Carvalhaes}}]{oas2014exploring}%
  \BibitemOpen
  \bibfield  {author} {\bibinfo {author} {\bibfnamefont {J.}~\bibnamefont
  {Oas}}, \bibinfo {author} {\bibfnamefont {J.~A.}\ \bibnamefont {de~Barros}},\
  and\ \bibinfo {author} {\bibfnamefont {C.}~\bibnamefont {Carvalhaes}},\
  }\href {https://doi.org/10.1088/0031-8949/2014/t163/014034} {\bibfield
  {journal} {\bibinfo  {journal} {Phys. Scr.}\ }\textbf {\bibinfo {volume}
  {T163}},\ \bibinfo {pages} {014034} (\bibinfo {year} {2014})}\BibitemShut
  {NoStop}%
\bibitem [{\citenamefont {Morris}\ \emph {et~al.}(2022)\citenamefont {Morris},
  \citenamefont {Fiderer}, \citenamefont {Lang},\ and\ \citenamefont
  {Goldwater}}]{morris2022witnessing}%
  \BibitemOpen
  \bibfield  {author} {\bibinfo {author} {\bibfnamefont {B.}~\bibnamefont
  {Morris}}, \bibinfo {author} {\bibfnamefont {L.~J.}\ \bibnamefont {Fiderer}},
  \bibinfo {author} {\bibfnamefont {B.}~\bibnamefont {Lang}},\ and\ \bibinfo
  {author} {\bibfnamefont {D.}~\bibnamefont {Goldwater}},\ }\href
  {https://doi.org/10.1103/PhysRevA.105.032202} {\bibfield  {journal} {\bibinfo
   {journal} {Phys. Rev. A}\ }\textbf {\bibinfo {volume} {105}},\ \bibinfo
  {pages} {032202} (\bibinfo {year} {2022})}\BibitemShut {NoStop}%
\bibitem [{\citenamefont {Onggadinata}\ \emph {et~al.}(2021)\citenamefont
  {Onggadinata}, \citenamefont {Kurzynski},\ and\ \citenamefont
  {Kaszlikowski}}]{onggadinata2021local}%
  \BibitemOpen
  \bibfield  {author} {\bibinfo {author} {\bibfnamefont {K.}~\bibnamefont
  {Onggadinata}}, \bibinfo {author} {\bibfnamefont {P.}~\bibnamefont
  {Kurzynski}},\ and\ \bibinfo {author} {\bibfnamefont {D.}~\bibnamefont
  {Kaszlikowski}},\ }\href@noop {} {\bibfield  {journal} {\bibinfo  {journal}
  {arXiv preprint arXiv:2106.07945}\ } (\bibinfo {year} {2021})}\BibitemShut
  {NoStop}%
\bibitem [{\citenamefont {Kochen}\ and\ \citenamefont
  {Specker}(1967)}]{kochen1967problem}%
  \BibitemOpen
  \bibfield  {author} {\bibinfo {author} {\bibfnamefont {S.}~\bibnamefont
  {Kochen}}\ and\ \bibinfo {author} {\bibfnamefont {E.~P.}\ \bibnamefont
  {Specker}},\ }\href {http://www.jstor.org/stable/24902153} {\bibfield
  {journal} {\bibinfo  {journal} {J. Math. Mech.}\ }\textbf {\bibinfo {volume}
  {17}},\ \bibinfo {pages} {59} (\bibinfo {year} {1967})}\BibitemShut {NoStop}%
\bibitem [{\citenamefont {Yu}\ and\ \citenamefont {Oh}(2012)}]{yu2012state}%
  \BibitemOpen
  \bibfield  {author} {\bibinfo {author} {\bibfnamefont {S.}~\bibnamefont
  {Yu}}\ and\ \bibinfo {author} {\bibfnamefont {C.~H.}\ \bibnamefont {Oh}},\
  }\href {https://doi.org/10.1103/PhysRevLett.108.030402} {\bibfield  {journal}
  {\bibinfo  {journal} {Phys. Rev. Lett.}\ }\textbf {\bibinfo {volume} {108}},\
  \bibinfo {pages} {030402} (\bibinfo {year} {2012})}\BibitemShut {NoStop}%
\bibitem [{\citenamefont {Specker}(1960)}]{specker1960logik}%
  \BibitemOpen
  \bibfield  {author} {\bibinfo {author} {\bibfnamefont {E.}~\bibnamefont
  {Specker}},\ }\href
  {https://doi.org/https://doi.org/10.1111/j.1746-8361.1960.tb00422.x}
  {\bibfield  {journal} {\bibinfo  {journal} {Dialectica}\ }\textbf {\bibinfo
  {volume} {14}},\ \bibinfo {pages} {239} (\bibinfo {year} {1960})},\ \Eprint
  {https://arxiv.org/abs/https://onlinelibrary.wiley.com/doi/pdf/10.1111/j.1746-8361.1960.tb00422.x}
  {https://onlinelibrary.wiley.com/doi/pdf/10.1111/j.1746-8361.1960.tb00422.x}
  \BibitemShut {NoStop}%
\bibitem [{\citenamefont {Cabello}\ \emph {et~al.}(1996)\citenamefont
  {Cabello}, \citenamefont {Estebaranz},\ and\ \citenamefont
  {Garc{\'\i}a-Alcaine}}]{cabello1996bell}%
  \BibitemOpen
  \bibfield  {author} {\bibinfo {author} {\bibfnamefont {A.}~\bibnamefont
  {Cabello}}, \bibinfo {author} {\bibfnamefont {J.}~\bibnamefont
  {Estebaranz}},\ and\ \bibinfo {author} {\bibfnamefont {G.}~\bibnamefont
  {Garc{\'\i}a-Alcaine}},\ }\href
  {https://doi.org/https://doi.org/10.1016/0375-9601(96)00134-X} {\bibfield
  {journal} {\bibinfo  {journal} {Phys. Lett. A}\ }\textbf {\bibinfo {volume}
  {212}},\ \bibinfo {pages} {183} (\bibinfo {year} {1996})}\BibitemShut
  {NoStop}%
\bibitem [{\citenamefont {Amaral}\ and\ \citenamefont
  {Cunha}(2018)}]{amaral2018graph}%
  \BibitemOpen
  \bibfield  {author} {\bibinfo {author} {\bibfnamefont {B.}~\bibnamefont
  {Amaral}}\ and\ \bibinfo {author} {\bibfnamefont {M.~T.}\ \bibnamefont
  {Cunha}},\ }\bibinfo {title} {On {G}raph {A}pproaches to {C}ontextuality and
  {T}heir {R}ole in {Q}uantum {T}heory}\ (\bibinfo  {publisher} {Springer},\
  \bibinfo {address} {Cham, Switzerland},\ \bibinfo {year} {2018})\BibitemShut
  {NoStop}%
\bibitem [{\citenamefont {Cabello}\ \emph {et~al.}(2014)\citenamefont
  {Cabello}, \citenamefont {Severini},\ and\ \citenamefont
  {Winter}}]{cabello2014graph}%
  \BibitemOpen
  \bibfield  {author} {\bibinfo {author} {\bibfnamefont {A.}~\bibnamefont
  {Cabello}}, \bibinfo {author} {\bibfnamefont {S.}~\bibnamefont {Severini}},\
  and\ \bibinfo {author} {\bibfnamefont {A.}~\bibnamefont {Winter}},\ }\href
  {https://doi.org/10.1103/PhysRevLett.112.040401} {\bibfield  {journal}
  {\bibinfo  {journal} {Phys. Rev. Lett.}\ }\textbf {\bibinfo {volume} {112}},\
  \bibinfo {pages} {040401} (\bibinfo {year} {2014})}\BibitemShut {NoStop}%
\bibitem [{\citenamefont {Ac{\'\i}n}\ \emph {et~al.}(2015)\citenamefont
  {Ac{\'\i}n}, \citenamefont {Fritz}, \citenamefont {Leverrier},\ and\
  \citenamefont {Sainz}}]{acin2015combinatorial}%
  \BibitemOpen
  \bibfield  {author} {\bibinfo {author} {\bibfnamefont {A.}~\bibnamefont
  {Ac{\'\i}n}}, \bibinfo {author} {\bibfnamefont {T.}~\bibnamefont {Fritz}},
  \bibinfo {author} {\bibfnamefont {A.}~\bibnamefont {Leverrier}},\ and\
  \bibinfo {author} {\bibfnamefont {A.~B.}\ \bibnamefont {Sainz}},\ }\href
  {https://doi.org/https://doi.org/10.1007/s00220-014-2260-1} {\bibfield
  {journal} {\bibinfo  {journal} {Comm. Math. Phys.}\ }\textbf {\bibinfo
  {volume} {334}},\ \bibinfo {pages} {533} (\bibinfo {year}
  {2015})}\BibitemShut {NoStop}%
\bibitem [{\citenamefont {Gleason}(1957)}]{gleason1957measures}%
  \BibitemOpen
  \bibfield  {author} {\bibinfo {author} {\bibfnamefont {A.~M.}\ \bibnamefont
  {Gleason}},\ }\href {https://doi.org/https://www.jstor.org/stable/24900629}
  {\bibfield  {journal} {\bibinfo  {journal} {J. Math. Mech.}\ }\textbf
  {\bibinfo {volume} {6}},\ \bibinfo {pages} {885} (\bibinfo {year}
  {1957})}\BibitemShut {NoStop}%
\end{thebibliography}%

\end{document}